\documentclass[draftcls, onecolumn, 11pt]{IEEEtran} 

\usepackage {graphicx}  
\usepackage {amssymb}
\usepackage {amsmath}
\usepackage {mathrsfs}
\usepackage{amsmath}   

\begin{document}
%
\title{Adaptive Methods for Linear Programming Decoding}
\author{\authorblockN{Mohammad H. Taghavi N. and Paul H. Siegel}\\
\authorblockA{Department of Electrical and Computer Engineering\\
University of California, San Diego\\
Email: (mtaghavi, psiegel)@ucsd.edu}
}
%
%

\maketitle

\begin{abstract}
Detectability of failures of linear programming (LP) decoding and the
potential for improvement by adding new constraints motivate the use 
of an adaptive approach in selecting the constraints for the underlying 
LP problem.  In this paper, we make a first step in studying this method, 
and show that it can significantly reduce the complexity of the problem, 
which was originally exponential in the maximum check-node degree. 
We further show that adaptively adding new constraints, e.g. by 
combining parity checks, can provide large gains in the performance. 
\end{abstract}


%
\IEEEpeerreviewmaketitle

\newtheorem{definition}{Definition}
\newtheorem{theorem}{Theorem}
\newtheorem{lemma}{Lemma}
\newtheorem{claim}{Claim}
\newtheorem{algorithm}{Algorithm}
\newtheorem{corollary}{Corollary}
\newtheorem{conjecture}{Conjecture}
\newtheorem{remark}{Remark}
\section{Introduction}
Linear programming (LP) decoding, as an approximation to maximum-likelihood (ML) 
decoding, was proposed by Feldman \emph{et al.} \cite{Feldman thesis}. 
Many observations suggest similarities between the performance of LP and iterative 
message-passing decoding methods. For example, we know 
that the existence of low-weight pseudocodewords degrades the performance of both 
types of decoders \cite{Feldman thesis}, \cite{LP and MSA}, \cite{Vontobel}. Therefore, it is 
reasonable to try to exploit the simpler geometrical structure of LP decoding to 
make predictions about the performance of message-passing decoding algorithms.

On the other hand, there are differences between these two decoding approaches. For instance, given an LDPC code, we know 
that adding redundant parity checks that are satisfied by all the codewords can 
not degrade the performance of LP decoding, while with message-passing algorithms, these parity checks may have a negative effect by introducing short cycles in the corresponding Tanner graph. 
This property of LP decoding allows performance improvement by tightening the relaxation. 
Another characteristic of LP decoding -- the \emph{ML certificate property} -- is that 
its failure to find an ML codeword is always detectable. More specifically, the 
decoder always gives either an ML codeword or a nonintegral pseudocodeword as the solution. 

These two properties motivate the use of an adaptive approach in LP decoding which can be summarized as follows: Given a set of constraints that describe a code, start the LP decoding with a few of them, then sequentially and adaptively add more of the constraints to the problem until either an ML codeword is found or no further ``useful'' constraint exists. The goal of this paper is to explore the potential of this idea for improving the performance of LP decoding. 

We show that by incorporating adaptivity into the LP decoding procedure,
we can achieve with a small number of constraints an error-rate performance comparable to that obtained when standard LP decoding is applied to a relaxation defined by a much larger number of constraints. 
In particular, we observe that while the number of constraints per check node required for convergence of LP decoding is exponential in the check node degrees, the adaptive method generally converges with a (small) constant number of constraints that does not appear to be dependent upon the underlying code's degree distribution. This property makes it feasible to apply LP decoding to higher-density graph codes. 

Along the way, we prove several general properties of LP relaxations of ML decoding that shed light upon the performance of LP and iterative decoding algorithms.

The rest of this paper is organized as follows. 
In Section II, we review Feldman's LP decoding. 
In Section III, we introduce and analyze an adaptive algorithm to solve the original LP problem more efficiently. 
In Section IV, we study how adaptively imposing additional constraints can improve the LP decoder performance. 
Section V concludes the paper. 
%

\section{LP Relaxation of ML Decoding}
Consider a binary linear code $\mathscr{C}$ of length $n$. If a codeword $\underline 
y\in \mathscr{C}$ is transmitted through a memoryless binary-input output-symmetric (MBIOS) 
channel, the ML codeword given the received vector $\underline r \in \mathbb{R}^n$ is 
the solution to the optimization problem
\begin{eqnarray}
\label{ML decoding}
\textrm{minimize} &&\underline \gamma^T \underline x \nonumber\\
\textrm{subject to} &&\underline x\in \mathscr{C},
\end{eqnarray}
where $\underline \gamma$ is the vector of log-likelihood ratios defined as
\begin{equation}
\label{def gamma}
\gamma_i= \log \left(\frac{\Pr(r_i|y_i=0)}{\Pr(r_i|y_i=1)} \right).
\end{equation}
As an approximation to ML decoding, Feldman \emph{et al.} proposed a relaxed version of 
this problem by first considering the convex hull of the local codewords defined by each 
row of the parity-check matrix, and then intersecting them to obtain what is called the 
\emph{fundamental polytope}, $\mathscr{P}$, by Koetter \emph{et al.} \cite{Vontobel}. 
This polytope has a number of integral and nonintegral vertices, but the integral vertices 
exactly correspond to the codewords of $\mathscr{C}$. Therefore, whenever LP decoding gives an 
integral solution, it is guaranteed to be an ML codeword. 

In Feldman's relaxation of the decoding problem, constraints are derived from a parity-check matrix as follows. For each row $j=1,\ldots,m$ of the parity-check matrix, define the neighborhood set $N(j) \subset \{1, 2, \ldots, n\}$ of the corresponding check node in the Tanner graph to be the variable nodes that are directly connected to it. (For convenience, we often identify check nodes and variable nodes with their respective index sets $j=1,\ldots,m$ and $i=1,\ldots,n$.) Then, for $j=1,\ldots,m$ the LP relaxation includes all of the following constraints:
\begin{equation}
\label{constraints}
\sum_{i\in V} x_i -\sum_{i\in N(j)\backslash V} x_i \leq |V|-1,\ \ \forall\ V\subset N(j)\ 
\textrm{such that}\ |V|\ \textrm{is odd}.
\end{equation}
Throughout the paper, we refer to the constraints of this form as \emph{parity-check constraints}. 
In addition, for any element $x_i$ of the optimization variable, $\underline x$, the constraint 
that $0\leq x_i \leq 1$ is also added.

\section{Adaptive LP Decoding}
As any odd-sized subset $V$ of the neighborhood $N$ of each check node introduces a unique 
parity-check constraint, there are $2^{d_c-1}$ constraints corresponding to each check node of 
degree $d_c$. Therefore, the total number of constraints and hence, the complexity of the problem, 
is exponential in terms of the maximum check node degree, $d_c^{max}$. This becomes more 
significant in a high density code where $d_c^{max}$ increases with the code length, $n$. 
In this section, we show that Feldman's LP relaxation has some properties that allow 
us to solve the optimization problem by using a much smaller number of constraints.

\subsection{Properties of the Relaxation Constraints}
\begin{definition}
Given a constraint of the form
\begin{equation}
\label{general const}
\underline a_i^T \underline x\leq b_i,
\end{equation}
and a vector $\underline x_0 \in \mathbb{R}^n$, we call (\ref{general const}) an 
\emph{active constraint} at $\underline x_0$ if
\begin{equation}
\label{active const}
\underline a_i^T \underline x_0=b_i,
\end{equation}
and a \emph{violated constraint} or, equivalently, a \emph{cut} at $\underline x_0$ if
\begin{equation}
\label{violated const}
\underline a_i^T \underline x_0>b_i.
\end{equation}
\end{definition}

A constraint that generates a cut at point $\underline x$ corresponds to a subset $V \subset N$ of odd cardinality such that 
\begin{equation}
\label{typical cut}
\sum_{i\in V} x_i -\sum_{i\in N\backslash V} x_i > |V|-1.
\end{equation}
This condition implies that 
\begin{equation}
\label{obsv1}
|V|-1 < \sum_{i\in V} x_i \leq |V|
\end{equation}
and
\begin{equation}
\label{obsv2}
0\leq \sum_{i\in N\backslash V}\!\!\! x_i < x_{\ell}, \ \forall \ell\in V.
\end{equation}
The following theorem reveals a special property of the constraints of 
the LP decoding problem.
\begin{theorem}
\label{one cut per check}
At any given point $\underline x \in [0,1]^n$, at most one of the constraints 
introduced by each check node can be a cut.
\end{theorem}
\begin{proof}
Consider a check node with neighborhood $N\subset \{1, 2, \ldots , n\}$ and 
two subsets $V_1\subset N$ and $V_2\subset N$ of odd sizes $\left|V_1\right|$ 
and $\left| V_2 \right|$, respectively, that each introduce a cut at point 
$\underline x$. We prove the theorem by showing that these two cuts must 
be identical, i.e. $V_1 = V_2$. 

Partition $N$ into four disjoint subsets $S=V_1\cap V_2$, $\overline 
V_1=V_1\backslash V_2$, $\overline V_2 = V_2\backslash V_1$, and 
$\overline N = N\backslash (V_1 \cup V_2)$. 
Now we can write the two corresponding constraints as
\begin{equation}
\label{cut1}
\sum_{i\in S} x_i +\sum_{i\in \overline V_1} x_i -\sum_{i\in \overline V_2} x_i - 
\sum_{i\in \overline N} x_i > |S|+\left| \overline V_1 \right|-1,
\end{equation}
and
\begin{equation}
\label{cut2}
\sum_{i\in S} x_i +\sum_{i\in \overline V_2} x_i -\sum_{i\in \overline V_1} x_i - 
\sum_{i\in \overline N} x_i > |S|+\left| \overline V_2 \right|-1.
\end{equation}
Now, we add the two inequalities and divide both sides by 2 to get
\begin{equation}
\label{cut1+2}
\sum_{i\in S} x_i - \sum_{i\in \overline N} x_i > |S| + \frac{\left| \overline V_1 
\right| + \left| \overline V_2 \right|}{2}-1.
\end{equation}
Since $x_i\in [0, 1]$ for every $i$, the left-hand side is less than or equal to 
$|S|$. Hence, for the right-hand side, we should have
\begin{equation}
|S|+\frac{\left| \overline V_1 \right| + \left| \overline V_2 \right|}{2}-1 < |S|,
\end{equation}
which yields
\begin{equation}
\label{sum of sizes}
\left| \overline V_1 \right| + \left| \overline V_2 \right| <2.
\end{equation}
Knowing that $\left|V_1\right|=\left|S\right|+\left|\overline V_1\right|$ and 
$\left|V_2\right|=\left|S\right|+\left|\overline V_2\right|$ are both positive 
odd numbers, we conclude that their difference, $\left|V_1\right|-\left|V_2\right| 
= \left|\overline V_1\right|-\left|\overline V_2\right|$ is an even number. 
Therefore $\left|\overline V_1\right|+\left|\overline V_2\right|$ is an even 
number, as well. Hence, (\ref{sum of sizes}) can hold only if $\left|\overline 
V_1\right|+\left|\overline V_2\right|=0$, which means that $\left|\overline 
V_1\right|=\left|\overline V_2\right|=0$. It follows that $V_1$ and $V_2$ are 
identical.
\end{proof}

Given an $(n, k)$ linear code with $m=n-k$ parity checks, a natural question is 
how we can find all the cuts defined by the LP relaxation at any given point 
$\underline x \in \mathbb{R}^n$. 
Referring to (\ref{obsv2}), we see that for any check node and any odd-sized subset $V$ of its neighborhood $N$ that introduces a cut, the variable nodes in $V$ have the largest values among all of the nodes in $N$.
Therefore, sorting the elements of $\underline x$ can simplify the process of   searching for a cut.  This observation is reflected in Algorithm~\ref{find cuts} below.

Consider a check node with neighborhood $N$. Without loss of generality, assume that variable nodes in $N$ have indices $1, 2, \ldots , |N|$,  and that their values satisfy $x_1\geq x_2 \geq \cdots \geq x_{|N|}$. The following algorithm provides an efficient way to find the unique cut generated by this check node at $\underline{x}$, if a cut exists.

\begin{algorithm}
\label{find cuts}
\end{algorithm}

\emph{Step 1: Set $v=1$, $V=\{1\}$ and $V^c\triangleq N\backslash V=\{2, 3, \ldots, 
|N|\}$.}

\emph{Step 2: Check the constraint (\ref{constraints}). If it is violated, 
we have found the cut. Exit.}

\emph{Step 3: Set $v=v+2$. If $v\leq |N|$, move $x_{v-1}$ and $x_{v}$ (the two 
largest members of $V^c$) from $V^c$ to $V$}

\emph{Step 4: If $v\leq |N|$ and (\ref{obsv1}) is satisfied, go to Step 2; 
otherwise, the check node does not provide a cut at $x$.}

Note that the failure of condition (\ref{obsv1}) provides a definitive termination criterion for the algorithm when no cut exists.
If redundant calculations are avoided in calculating the sums in (\ref{constraints}), this algorithm can find the cut generated 
by the check node, if it exists, in $O(d_c)$ time, where $d_c=|N|$ is the degree 
of the check node. 
Repeating the procedure for each check node, and considering $O(n\log n)$ 
complexity for sorting $\underline x$, the time required to find all the cuts at 
point $\underline x$ becomes $O(m d_c^{max} + n\log n)$ 
\footnote[1]{For low-density parity-check codes, it is better to sort the neighbors of each 
check node separately, so the total complexity becomes $O(m d_c^{max} + 
m d_c^{max}\log d_c^{max})$.}
.

\subsection{The Adaptive Procedure}
The fundamental polytope for a parity-check code is defined by a large number of 
constraints (hyperplanes), and a linear programming solver finds the vertex of 
this polytope that minimizes the objective function, or, in other words, the 
pseudocodeword that is closest to the received vector. For example, the Simplex 
algorithm starts from an initial vertex and visits different vertices of the polytope by 
traveling along the edges, until it finds the optimum vertex. The time required 
to find the solution is approximately proportional to the number of vertices that 
have been visited, and this, in turn, is determined by the number and properties 
of the constraints in the problem. Hence, if we eliminate some of the intermediate 
vertices and only keep those which are close to the optimum point, we can reduce 
the complexity of the algorithm. To implement this idea in the adaptive LP decoding 
scheme, we run the LP solver with a minimal number of constraints to ensure 
boundedness of the solution, and depending on the LP solution, we add only the 
``useful constraints'' that cut the current solution from the feasible region. 
This procedure is repeated until no further cut exists, which means that the 
solution is a vertex on the fundamental polytope.

To start the procedure, we need at least $n$ constraints to determine  
a vertex that can become the solution of the first iteration. Recalling that $0\leq x_i \leq 1$, we add for each $i$ exactly one of the constraints implied by these bounds. The choice depends upon whether increasing $x_i$ leads to an increase or decrease in the objective function. Specifically, for each $i\in \{1, 2, \ldots, n\}$, we introduce the initial constraint
\begin{align}
\label{initial constraints}
0\leq x_i \ \ &\text{if}\ \gamma_i>0, \nonumber \\
&\text{or} \nonumber\\
x_i\leq 1 \ \ &\text{if}\ \gamma_i<0.
\end{align}
Note that the optimum (and only) vertex satisfying this initial problem corresponds to the result of an (uncoded) bit-wise, hard decision based on the received vector. 

The following algorithm describes the adaptive LP decoding procedure.

\begin{algorithm}
\label{adaptive LP}
\end{algorithm}

\emph{Step 1: Setup the initial problem according to (\ref{initial constraints}).}

\emph{Step 2: Run the LP solver.}

\emph{Step 3: Find all cuts for the current solution.}

\emph{Step 4: If one or more cuts are found, add them to the problem constraints 
and go to Step 2. If not, we have found the solution. Exit.}

\medskip

\begin{lemma}
If no cut is found after any iteration of Algorithm \ref{adaptive LP}, the current solution represents the solution of the LP decoding problem incorporating all of the relaxation constraints given in Section II.
\end{lemma}
\begin{proof}
At any intermediate step of the algorithm, the space of feasible points with 
respect to the current constraints contains the fundamental polytope $\mathscr{P}$, as these constraints are all among the original constraints used to define $\mathscr{P}$. If at any iteration, no cut is found, we conclude that all the original constraints given by (\ref{constraints}) are satisfied by the current solution, $\underline x$, which means that this point is in $\mathscr{P}$. Hence, since $\underline x$ has the lowest 
cost in a space that contains $\mathscr{P}$, it is also the optimum 
point in $\mathscr{P}$. 
\end{proof}

To further speed up the algorithm, we can use a ``warm start'' after adding a 
number of constraints at each iteration. In other words, since the intermediate 
solutions of the adaptive algorithm converge to the solution of the original LP 
problem, we can use the solution of each iteration as a starting point for the 
next iteration. Since the initial point will, in principle, be close to the next solution, the number of steps of the Simplex algorithm, and therefore, the overall running time, is expected to decrease. 
On the other hand, each of these warm starts represents an infeasible point for the subsequent problem, since it will not satisfy the new constraints. As a result, the LP solver will have to first take a number of steps to move back into the feasible region. 
In Subsection D, we will discuss in more detail the effect of using warm starts on the speed of the algorithm.

\subsection{A Bound on the Complexity}
\begin{theorem}
\label{n iterations}
The adaptive algorithm (Algorithm \ref{adaptive LP}) converges after at 
most $n$ iterations.
\end{theorem}
\begin{proof}
The solution produced by the algorithm is a vertex $\underline x_f$ of the problem space determined 
by the initial constraints along with those added by the successive iterations of the cut-finding procedure. 
Therefore, we can find $n$ such constraints
$$
\kappa_i:\ \underline \alpha_i^T \underline x \leq \beta_i,\ i=1, 2,\ldots n,
$$
whose corresponding hyperplanes uniquely determine this vertex. 
This means that if we set up an LP problem with 
only those $n$ constraints, the optimal point will be $\underline x_f$. 
Now, consider the $k$th intermediate solution, $\underline x_k$, that is eliminated 
at the end of the $k$th iteration. At least one of the constraints, 
$\kappa_1, \ldots, \kappa_n$, should be violated by $\underline x_k$;
otherwise, since $\underline x_k$ has a lower cost than $\underline x_f$, 
$\underline x_k$ would be the solution of LP with these $n$ constraints. 
But we know that the cuts added at the $k$th iteration are all the possible 
constraints that are violated at $\underline x_k$. Consequently, at least one 
of the cuts added at each iteration should be among $\{\kappa_i: i=1, 2, 
\ldots,n\}$; hence, the number of iterations is at most $n$.
\end{proof}
\begin{remark}
The adaptive procedure and convergence result can be 
generalized to any LP problem defined by a fixed set of constraints.  
In general, however, there may not be an analog of 
Theorem~\ref{one cut per check} to facilitate the search for cut 
constraints.
\end{remark}

\begin{remark}
\label{int positions in PCW}
If for a given code of length $n$, the adaptive algorithm converges with at 
most $q<n$ final parity-check constraints, then each pseudocodeword of this 
LP relaxation should have at least $n-q$ integer elements. To see this, note 
that each pseudocodeword corresponds to the intersection of at least $n$ 
active constraints. 
If the problem has at most $q$ parity-check constraints, then at least $n-q$ 
constraints of the form $x_i\geq 0$ or $x_i\leq 1$ should be active at each 
pseudocodeword, which means that at least $n-q$ positions of the 
pseudocodeword are integer-valued.
\end{remark}

\begin{corollary}
The final application of the LP solver in the adaptive decoding algorithm 
uses at most $n(m+1)$ constraints.
\end{corollary}
\begin{proof}
The algorithm starts with $n$ constraints, and according to 
Theorem~\ref{one cut per check}, 
at each iteration no more than $m$ new constraints are added. 
Since there are at most $n$ 
iterations, the final number of constraints is less than or equal to $n(m+1)$.
\end{proof}

For high-density codes of fixed rate, this bound guarantees convergence with 
$O(n^2)$ constraints, whereas the standard LP relaxation requires a number 
of constraints that is exponential in $n$, and the high-density polytope 
representation given in \cite[Appendix II]{Feldman thesis} involves 
$O(n^3)$ variables and constraints. 

\subsection{Numerical Results}
To empirically investigate the complexity reduction due to the adaptive 
approach for LP decoding, 
we performed simulations over random regular LDPC codes of various lengths, 
degrees, and rates on the AWGN channel. 
All the experiments were performed with 
the low SNR value of $-1.0$ dB, since in the high SNR regime the 
received vector 
is likely to be close to a codeword, in which case the algorithm converges 
fast, rather than demonstrating its worst-case behavior. 

\begin{figure}
\centering
\includegraphics[width=3.5 in] {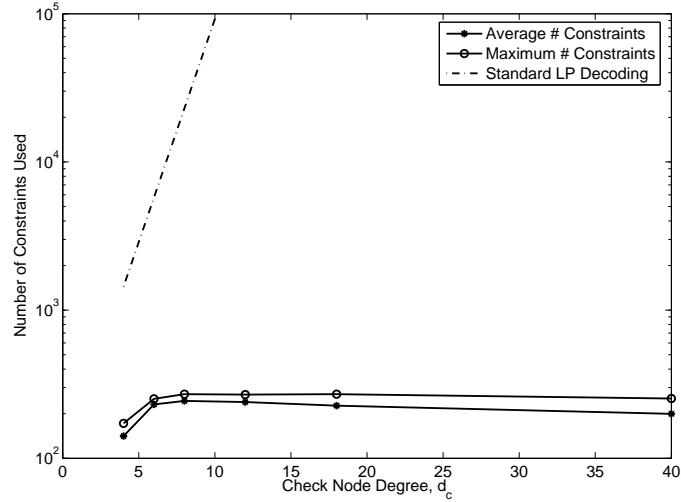}
\caption{The average and maximum number of parity-check constraints used 
versus check node degree, $d_c$, for fixed length $n=360$ and rate 
$R=\frac{1}{2}$.}\!\!\!\!\!\!\!\!
\label{const vs d_c}
\end{figure}

In the first scenario, we studied the effect of changing the check node degree 
$d_c$ from $4$ to $40$ while keeping the code length at $n=360$ and the rate at 
$R=\frac{1}{2}$. The simulation was performed over 400 blocks for each value of 
$d_c$. The average (resp. maximum) number of iterations required to converge 
started from around $14.5$ (resp. $30$) for $d_c=4$, and decreased 
monotonically down to $5.9$ (resp. $9$) for $d_c=40$. The average and 
maximum numbers of parity-check constraints 
in the final iteration of the algorithm are plotted in Fig.~\ref{const vs d_c}. 
We see that both the average and the maximum values are almost constant, 
and remain below $270$ for all the values of $d_c$. For comparison, the total 
number of constraints required for the standard (non-adaptive) LP decoding 
problem, 
which is equal to $2^{d_c-1}$, is also included in this figure. 
The decrease in 
the number of required constraints translates to a large gain for the adaptive 
algorithm in terms of the running time. 

\begin{figure}
\centering
\includegraphics[width=3.5 in] {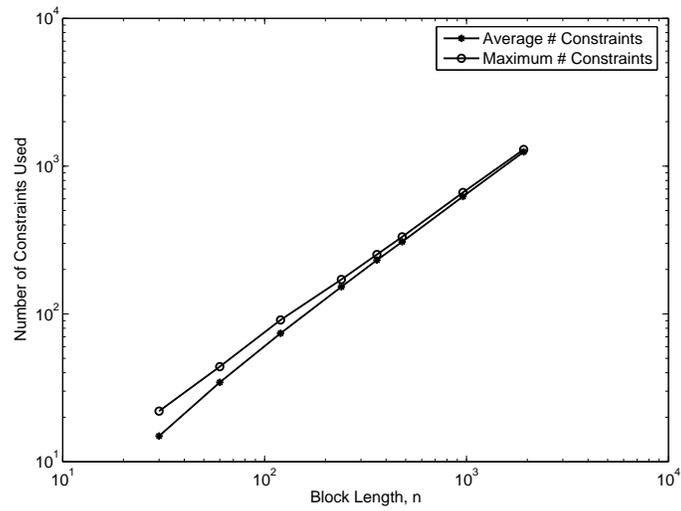}
\caption{The average and maximum number of parity-check constraints used 
versus block length, $n$, for fixed rate $R=\frac{1}{2}$ and check node 
degree $d_c=6$.}\!\!\!\!\!\!\!\!
\label{const vs n}
\end{figure}

In the second case, we studied random (3,6) codes of lengths $n=30$ to 
$n=1920$. 
For all values of $n$, the average (resp. maximum) number of required 
iterations 
remained between $5$ and $11$ (resp. $10$ and $16$). 
The average and maximum numbers of parity-check constraints in the final 
iteration are plotted versus $n$ in Fig.~\ref{const vs n}. 
We observe that the number of constraints is generally between 
$0.6 n$ and $0.7 n$. 

\begin{figure}
\centering
\includegraphics[width=3.5 in] {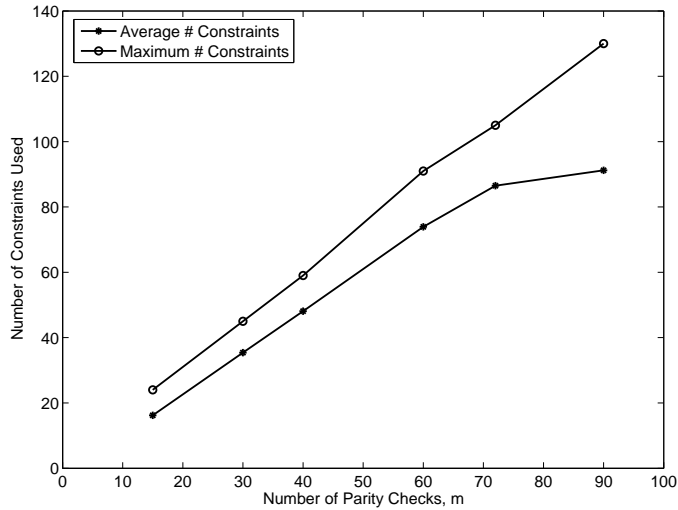}
\caption{The average and maximum number of parity-check constraints used 
versus the number of parity checks, $m$, for $n=120$ and $d_v=3$.}\!\!\!\!\!\!\!\!
\label{const vs m}
\end{figure}

In the third experiment, we investigated the effect of the rate of the code 
on the performance of the algorithm. Fig.~\ref{const vs m} shows the average  
and maximum numbers of parity-check constraints in the final iteration where 
the block length is $n=120$ and the number of  parity checks, $m$, 
increases from $15$ to $90$. 
The variable node degree is fixed at $d_v=3$. We see that the average and 
maximum numbers of constraints are respectively in the ranges $1.1m$ to 
$1.2m$ and $1.4m$ to $1.6m$ for most values of $m$. The relatively large 
drop in the 
average number for $m=90$ with respect to the linear curve can be explained 
by the fact that at this value of $m$ the rate of failure of LP decoding was 
less than $0.5$ at $-1.0$ dB, whereas for all the other values of $m$, this 
rate was close to $1$. Since the success of LP decoding generally indicates 
proximity of the received vector to a codeword, we expect the number of 
parity checks required to converge to be small in such a case, which 
decreases the average number of constraints.

\begin{figure}
\centering
\includegraphics[width=3.5 in] {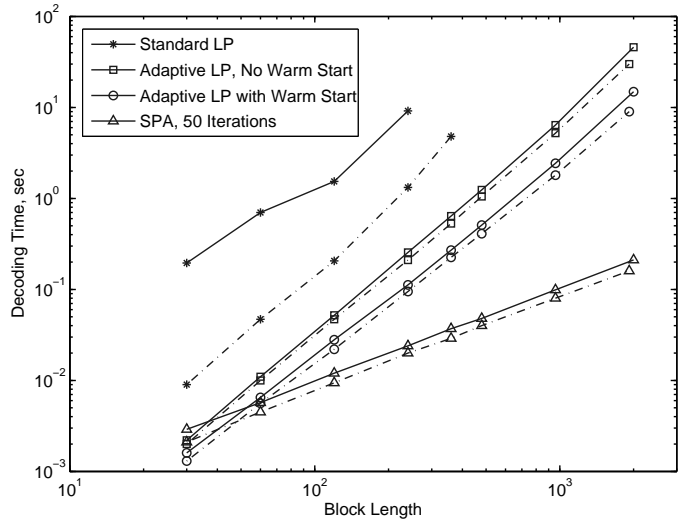}
\caption{The average decoding time versus the length of the code for regular 
(3,6) LDPC codes (dashed lines) and (4,8) LDPC codes (solid lines) at 
SNR=$-1.0$ dB.}\!\!\!\!\!\!\!\!
\label{time vs n}
\end{figure}

Finally, in Fig.~\ref{time vs n}, we compare the average decoding time of 
different algorithms at the low SNR of $-1.0$ dB. It is important to note 
that the running times of the LP-based techniques strongly depend on the 
underlying LP solver. In this work, we have used the open-source GNU Linear 
Programming Kit (GLPK \cite{GLPK}) for solving the LPs. 
The numerical results demonstrate that the adaptive algorithm significantly 
reduces the gap between the speed of standard LP decoding and that of the 
sum-product message-passing algorithm. 
Comparing the results for the (3,6) codes (dashed lines) and the (4,8) codes 
(solid lines) further shows that while the decoding time for the standard LP 
increases very rapidly with the check node degree of the code, the adaptive 
technique is not significantly affected by the change in the check node degree. 

Our simulations indicate that the decoding time of the adaptive algorithm 
does not change significantly even if the code has a high-density 
parity-check matrix. 
This result can be explained by two factors. First, Fig.~\ref{const vs d_c} 
shows that the number of constraints used in the algorithm 
does not change with the check node degree of the code. 
Second, while having a smaller check degree makes the matrix of constraints 
sparser, the LP solver  that we are using does not benefit from this sparsity. 
A similar behavior was also observed when we used a commercial LP solver, 
MOSEK, instead of GLPK.
We expect that by designing a special LP solver than can effectively take 
advantage of the sparsity of this problem, the time 
complexities of the LP-based techniques may become closer to those of the 
message-passing techniques.

Fig.~\ref{time vs n} also shows the average decoding time when warm starts 
are used in the iterations of the adaptive decoding algorithm. 
We can see that warm starts slightly decrease the slope of the decoding-time 
curve when plotted against the logarithm of the block length.  
This translates into approximately a factor of 3 improvement in the 
decoding time at a block length of 1000.

Based on the simulation results, we observe that in practice the algorithm 
performs much faster than is guaranteed by Theorem \ref{n iterations}. 
These observations suggest the following conjecture. 
\begin{conjecture}
\label{conjecture on complexity}
For a random parity-check code of length $n$ with $m=n(1-R)$ parity checks 
and arbitrary degree distributions, as $n$ and $m$ increase, the adaptive 
LP decoding algorithm converges with probability arbitrarily close to $1$ 
in at most $\alpha$ iterations and with at most an average of $\beta$ final 
parity-check constraints per check node, where $\alpha$ and $\beta$ are 
constants independent of the length, rate and degree distribution of the code.
\end{conjecture}

\section{Generating Cuts to Improve the Performance}
The complexity reduction obtained by adaptive LP decoding inspires the use of 
cutting-plane techniques to improve the error rate performance of the 
algorithm. 
Specifically, when LP with all the original constraints gives a nonintegral 
solution, we try to cut the current solution, while keeping all the possible 
integral solutions (codewords) feasible.  

In the decoding problem, the new cuts can be chosen from a pool of constraints 
describing a relaxation of the maximum-likelihood problem which is tighter than 
the fundamental polytope. 
In this sense, the cutting-plane technique is equivalent 
to the adaptive LP decoding of the previous section, with the difference that 
there are more constraints to choose from. The effectiveness of this method 
depends on how closely the new relaxation approximates the ML decoding problem, 
and how efficiently we can search for those constraints that introduce cuts. 
Feldman \emph{et al.} \cite{Feldman thesis} have mentioned some ways to tighten 
the relaxation of the ML decoding, including adding redundant parity checks 
(RPC), 
and using lift-and-project methods. (For more on lift-and-project, 
see \cite{lift-and-project} and references therein.) Gomory's algorithm 
\cite{Gomory} is also one of the most well-known techniques for general 
integer optimization problems, although it suffers from slow convergence. 
Each of these methods can be applied adaptively in the context of 
cutting-plane techniques.

The simple structure of RPCs makes them an interesting choice for 
generating cuts. 
There are examples, such as the dual code of the (4,7) Hamming code, 
where even the relaxation obtained by adding all the possible 
RPC constraints does not guarantee convergence to a codeword. In other words, 
it is possible to obtain a nonintegral solution for which there is no RPC 
cut. Understanding the effect of RPCs in general requires further study. 
Also, finding efficient methods to search for RPC cuts for a given 
nonintegral solution remains an open issue. 
On the other hand, as observed in simulation results, RPC cuts are 
generally strong, and a reasonable number of them makes the resulting 
LP relaxation tight enough to converge to an integer-valued 
solution. In this work, we focus on cutting-plane algorithms that use RPC cuts.

\subsection{Finding Redundant Parity-Check Cuts}
An  RPC is obtained by modulo-2 addition of some of the rows 
of the parity-check matrix, and this new check introduces a number of 
constraints that may include a cut. There is an exponential number of RPCs 
that can be made this way, and in general, most of them do not introduce cuts. 
Hence, we need to find the cuts efficiently by exploiting the particular 
structure of the decoding problem. In particular, we observe that cycles 
in the graph have an important role in determining whether an RPC 
generates 
a cut. To explain this property, we start with some definitions. 

\begin{definition}
\label{minimal subset}
Given a current solution, $\underline x$, the subset $T \subset \{1, 2, 
\ldots, m\}$ of check nodes is called a \emph{cut-generating collection} 
if the RPC made by modulo-2 addition of the parity-checks corresponding to 
$T$ introduces a cut. If no proper of $T$ other that itself has this property, 
we call it a \emph{minimal cut-generating collection}.
\end{definition}

\begin{definition}
\label{fractional subgraph}
Given a pseudo-codeword $\underline x$, we denote by $\phi$ the set of 
variable nodes in the Tanner graph of the code whose corresponding elements 
in $\underline x$ have fractional values. Also, let $F$ be the subgraph made 
up of these variable nodes, the check nodes directly connected to them, and 
all the edges that connect them. We call $F$ the \emph{fractional subgraph} 
and any cycle in $F$ a \emph{fractional 
cycle} at $\underline x$.
\end{definition}

Theorem~\ref{RPC cycles} below explains the relevance of the concept 
of fractional cycles. Its proof makes use of the following lemma.
\begin{lemma}
\label{two edges}
Suppose that $c_1$ and $c_2$ are two parity checks whose constraints are
satisfied by the current solution, $\underline x$.
Then, $c\triangleq c_1 \oplus c_2$, the modulo-2 combination of these
checks, can generate a cut only if the neighborhoods of $c_1$ and $c_2$
have at least two fractional-valued variable nodes in common.
\end{lemma}
\begin{proof}
See Appendix I.
\end{proof}

\begin{theorem}
\label{RPC cycles}
Let $T$ be a collection of check nodes in the Tanner graph of the code. 
If $T$ is a cut-generating collection at $\underline x$, then there exists 
a fractional cycle such that all the check nodes on it belong to $T$.
\end{theorem}
\begin{proof}
We first consider the case where $T$ is a minimal cut-generating collection.
Note that any cut-generating collection must contain at least two check 
nodes, since no single check node generates a cut.  
Pick an arbitrary check node $c_j$ in $T$. 
We make an RPC $c_r$ by linearly combining the parity checks in 
$T\backslash\{c_j\}$. 
From Lemma~\ref{two edges} and the minimality of $T$, it follows that 
there are at least two fractional-valued variable nodes common to the 
neighborhoods of $c_j$ and $c_r$.  
Applying this reasoning to every check node in $T$, we conclude that 
any check node in the collection $T$ is connected to at least two 
fractional-valued variable nodes of degree at least 2. 
Therefore, the subgraph $G$ corresponding to the check nodes in $T$ 
and their neighboring variable nodes must contain a cycle which passes 
through only fractional-valued variable nodes, as claimed.

If $T$ is not a minimal cut-generating collection, then it must contain 
a minimal cut-generating collection, $T_0$. 
To see this, observe that there must be a check node in $T$ whose removal 
leaves a cut-generating collection of check nodes. 
Iteration of this check node removal process must terminate in a 
non-empty minimal cut-generating collection $T_0$ containing at least 
two check nodes. 
The subgraph $G_0$ corresponding to $T_0$ is contained in the subgraph 
$G$ corresponding to $T$, so the fractional cycle in $G_0$ constructed 
as above is also a fractional cycle in $G$. 

\end{proof}

The following theorem confirms that a fractional cycle always exists
for any non-integer pseudocodeword.  
We can represent the fractional subgraph $F$
corresponding to the pseudocodeword $\underline x$
as a union of disjoint, connected subgraphs $\{F_i\}$, $i=1,\cdots,K$, for
some $K\geq 1$.
We refer to each connected subgraph $F_i$ as a \emph{cluster}. 

\begin{theorem}
\label{existence of cycle}
Let $\underline x$ be the solution of an LP decoding problem 
with Tanner graph $G$ and log-likelihood vector $\gamma$. 
Let $F$ denote the fractional subgraph corresponding to $\underline x$. 
Then each cluster $F_i$, $i=1,\cdots,K$ in $F$ contains a cycle.
\end{theorem}
\begin{proof}
See Appendix II.
\end{proof}

The results above motivate the following algorithm to search for 
RPC cuts.

\begin{algorithm}
\label{search for RPC cuts}
\end{algorithm}

\emph{Step 1: Given a solution $\underline x$, prune the Tanner 
graph by removing all the variable nodes with integer values.}

\emph{Step 2: Starting from an arbitrary check node, 
randomly walk through the pruned graph until a cycle 
is found.}

\emph{Step 3: Create an RPC by combining the rows of the 
parity-check matrix corresponding to the check nodes in 
the cycle.}

\emph{Step 4: If this RPC introduces a cut, add it to the 
Tanner graph and exit; otherwise go to Step 2.}

When the fractional subgraph contains many cycles and it is feasible 
to check only a small fraction of them, the randomized method described 
above can efficiently find cycles. However, when the cycles are few in 
number, this algorithm may actually check a number of cycles several 
times, while skipping some others. 
In this case, a structured search, such as one based on the depth-first 
search (DFS) technique, can be used to find all the simple cycles in the 
fractional subgraph. 
One can then check to see if any of them introduces an RPC cut. 
However, to guarantee that all the potential RPCs are checked, 
one will still need to modify this search to include complex cycles, 
as, in general, a complex cycle is more likely to generate an RPC cut 
than a simple cycle with a comparable number of edges.

As shown above, by exploiting some of the properties of the linear code 
LP decoding problem, one can expedite the search for RPC cuts.  
However, there remains a need for more efficient methods of finding RPC cuts.


\subsection{Complexity Considerations}
There are a number of parameters that determine the complexity 
of the adaptive algorithm with RPC cuts, including the number 
of iterations of Algorithm \ref{search for RPC cuts} to find 
a cut, the total number of cuts that are needed to obtain an 
integer solution, and the time taken by each run of the LP 
solver after adding a cut. 
In particular, we observe empirically that a number of cuts less 
than the length of the code is often enough to ensure 
convergence to the ML codeword. By using each solution of 
the LP as a warm start for the next iteration 
after adding further cuts, the time that each LP takes 
can be significantly reduced. For example, for a regular 
(3,4) code of length 100 with RPC cuts, although as many 
as 70 LP problems may have to be solved for a successful 
decoding, the total time that is spent on these LP problems 
is no more than 10 times that of solving the standard 
problem (with no RPC cuts). 
Moreover, if we allow more than one cut to be added 
per iteration, the number of these iterations can be 
further reduced.

Since Algorithm \ref{search for RPC cuts} involves a random 
search, there is no guarantee that it will find a cut 
(if one exists) in a finite number of iterations. 
In particular, we have observed cases where, even after 
a large number of iterations, no cut was found, while 
a number of RPCs were repeatedly visited. 
This could mean that either no RPC cut exists for these 
cases, or the cuts have a structure that makes them 
unlikely to be selected by our random search algorithm.
 
In order to control the complexity, we can impose a limit, 
$C^{max}$, on the number of iterations of the search, and 
if no cut is found after $C^{max}$ trials, we declare failure. 
By changing $C^{max}$, we can trade complexity with 
performance. Alternatively, we can put a limit, $T^{max}$, 
on the total time that is spent on the decoding process. 
In order to find a proper value for this maximum, we ran 
the algorithm with a very large value of $C^{max}$ and 
measured the total decoding time for the cases where the 
algorithm was successful in finding the ML solution. 
Based on these observations, we found that 10 times the 
worst-case running time of the adaptive LP decoding algorithm
of Section III serves as a suitable value for $T^{max}$.

\begin{figure}
\centering
\includegraphics[width=3.5 in] {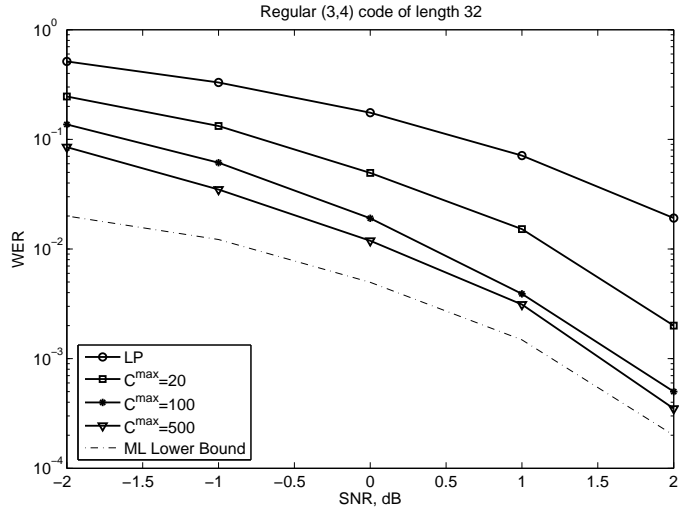}
\caption{WER of cutting-plane LP versus SNR for different 
values of $C^{max}$.}\!\!\!\!\!\!\!\!
\label{RPC WER}
\end{figure}

\subsection{Numerical Results}
To demonstrate the performance improvement achieved by 
using the RPC cutting-plane technique, we present simulation 
results for random regular $(3,4)$ LDPC codes on the AWGN 
channel. We consider codes of length 32, 100, and 240 bits.

In Fig.~\ref{RPC WER}, for the length-32 code, we plot the word 
error rate (WER) 
versus SNR for different values of $C^{max}$, demonstrating
the trade-off between performance and complexity. As in all 
subsequent figures, the SNR is defined as the ratio of the 
variance of the transmitted discrete-time signal to the 
variance of the noise sample.

For purposes of comparison, the WER of LP decoding with no 
RPC cut, as well as a lower bound on the WER of the ML decoder 
have been included in the figure. In order to obtain the ML 
lower bound, we counted the number of times that the cutting-plane 
LP algorithm, using a large value of $C^{max}$, converged to 
a codeword other than the transmitted codeword, and then divided 
that by the number of blocks. 
Due to the ML certificate property of LP decoding, we know 
that ML decoding would fail in those cases, as well. 
On the other hand, ML decoding may also fail in some of the 
cases where LP decoding does not converge to an integral solution. 
Therefore, this estimate gives a lower bound on the WER of 
ML decoding. 

However, this method for computing the ML lower bound could 
not be applied to the codes of length greater that 32 bits.
Therefore, as an alternative, we used the performance of the 
Box-and-Match soft decision decoding algorithm (BMA) developed 
by Valembois and Fossorier \cite{BMA} as an approximation
of the ML decoder performance.

\begin{figure}
\centering
\includegraphics[width=3.5 in] {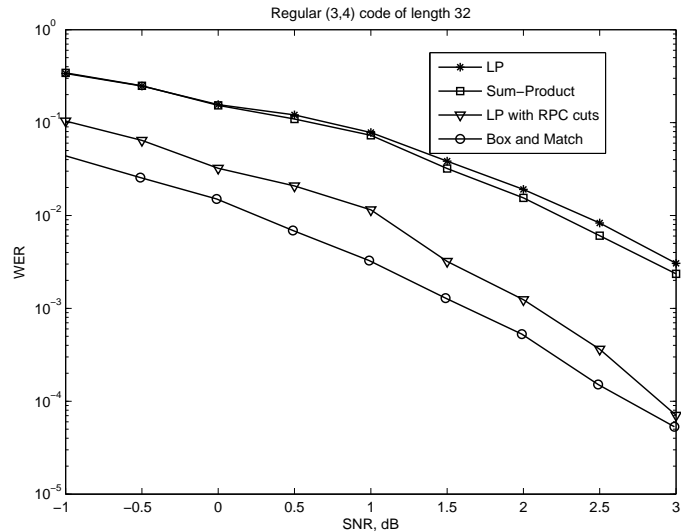}
\caption{WER of cutting-plane LP versus SNR for length 32 and
maximum decoding time 10 times that of the LP decoding.}\!\!\!\!\!\!\!\!
\label{RPC 32}
\end{figure}

\begin{figure}
\centering
\includegraphics[width=3.5 in] {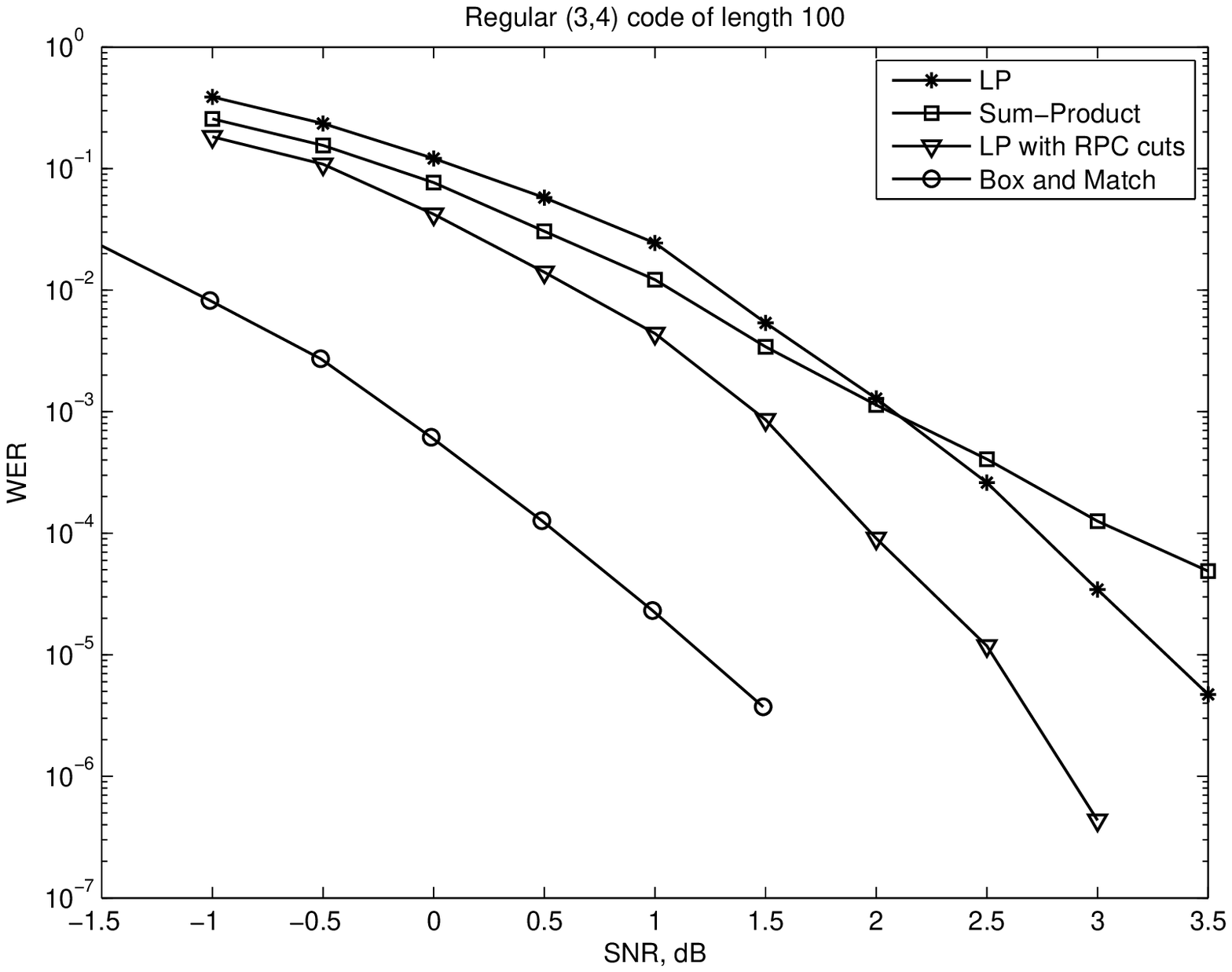}
\caption{WER of cutting-plane LP versus SNR for length 100 and 
maximum decoding time 10 times that of the LP decoding.}\!\!\!\!\!\!\!\!
\label{RPC 100}
\end{figure}

\begin{figure}
\centering
\includegraphics[width=3.5 in] {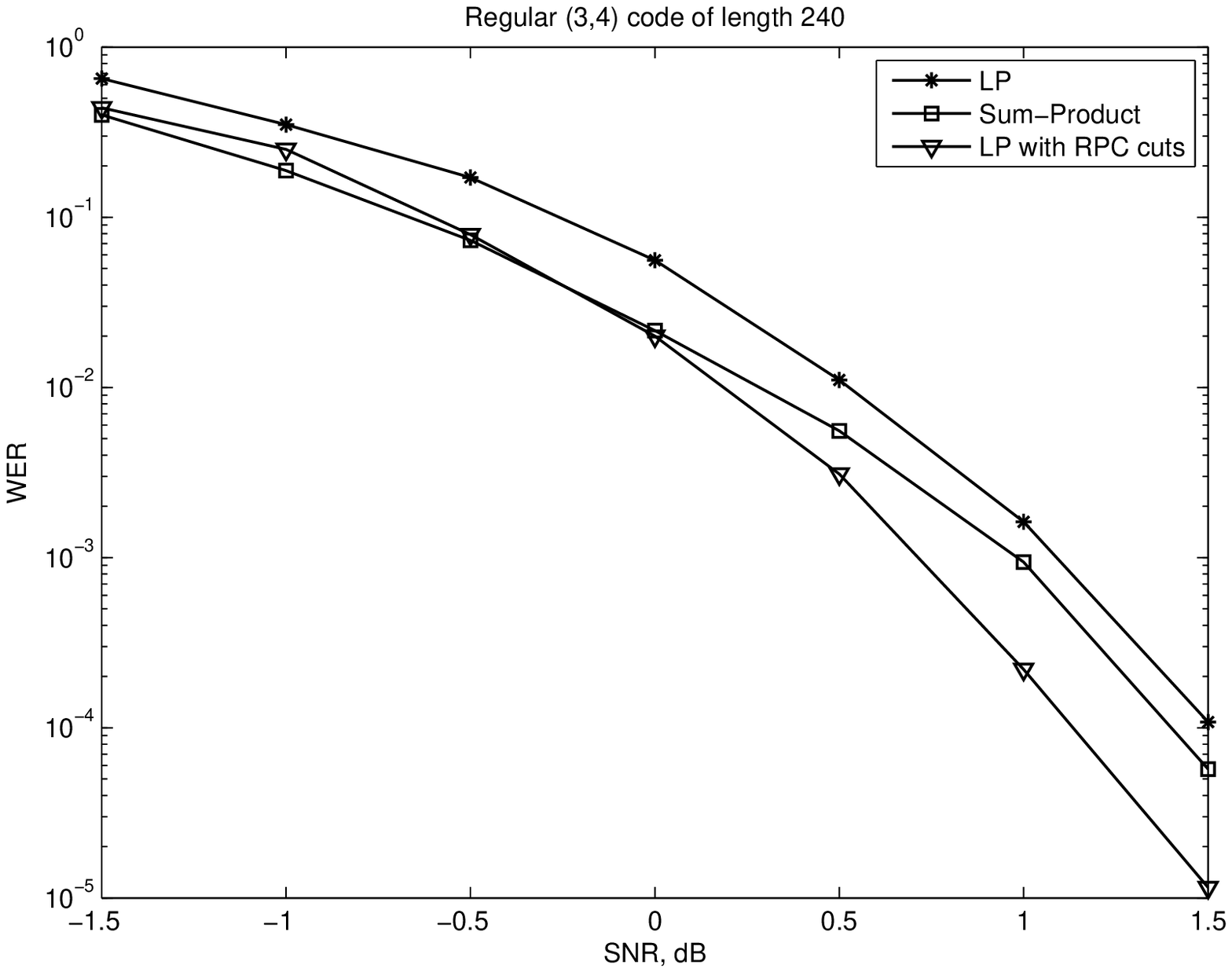}
\caption{WER of cutting-plane LP versus SNR for length 240 and 
maximum decoding time 10 times that of the LP decoding.}\!\!\!\!\!\!\!\!
\label{RPC 240}
\end{figure}

In Figs.~\ref{RPC 32}-\ref{RPC 240}, the performance of LP 
decoding with RPC cuts is compared to that of standard LP 
decoding, sum-product decoding, and also the BMA.
Each figure corresponds to a fixed block length, and in all three cases the sum-product decoding had 100 iterations.
The curves show that, as the SNR increases, 
the proposed method outperforms the LP and the SPA, and 
significantly closes the gap to the ML decoder performance.
However, one can see that, as the code length increases, the 
relative improvement provided by RPC cuts becomes less pronounced.
This may be due to the fact that, for larger code lengths, the 
Tanner graph becomes more tree-like, and therefore the negative 
effect of cycles on LP and message-passing decoding techniques 
becomes less important, especially at low SNR.

\section{Conclusion}
In this paper, we studied the potential for improving LP decoding, 
both in complexity and performance, by using an adaptive approach. 
Key to this approach is the ML certificate property of LP decoders,
that is, the ability to detect the failure to find the ML codeword.
This property is shared by message-passing decoding algorithms 
only in specific circumstances, such as on the erasure channel. 
The ML certificate property makes it possible to selectively  
and adaptively add only those constraints that are ``useful,'' 
depending on the current status of the LP decoding process.

We proposed an adaptive LP decoding algorithm with decoding 
complexity that is independent of the code degree distributions, 
making it possible to apply LP decoding to parity-check codes 
of arbitrary densities. 
However, since general purpose LP solvers are used at each iteration, 
the complexity is still a super-linear function of the block length,
as opposed to a linear function as achieved by the message-passing 
decoder. It remains an open question whether special LP solvers 
for decoding of LDPC codes might take advantage of the sparsity 
of the constraints and other properties of the LP decoding problem 
to provide linear complexity.

We also explored the application of cutting-plane techniques in 
LP decoding. 
We showed that redundant parity checks provide strong cuts, even 
though they may not guarantee ML performance. The results indicate
that it would be worthwhile to find more efficient ways to search for 
strong RPC cuts by exploiting their properties, as well as to 
determine specific classes of codes for which RPC cuts are 
particularly effective. It would also be interesting to investigate
the effectiveness of cuts generated by other techniques, 
such as lift-and-project cuts, Gomory cuts, or cuts specially 
designed for this decoding application. 
 
\appendices
\section{proof of Lemma \ref{two edges}}

\begin{proof}
Let $N_1$, $N_2$, and $N$ denote the sets of variable nodes in the 
neighborhoods of $c_1$, $c_2$, and $c$, respectively, and define 
$S\triangleq N_1\cap N_2$. Hence, we will have 
$N=(N_1\cup N_2)\backslash S$. We can write $S$ as a union of 
disjoint sets $S_0$, $S_1$, and $S_f$, where they respectively 
represent the members of $S$ whose values are equal to $0$, 
equal to $1$, or in the range $(0,1)$. In order to prove the 
lemma, it is enough to show that $|S_f|\geq 2$. 

Parity check $c$ generates a cut, hence there is an odd-sized 
subset $V\subset N$ of its neighborhood, for which we can write
\begin{equation}
\label{cut of c}
\sum_{i\in V} x_i - \sum_{i\in N\backslash V} x_i > |V|-1.
\end{equation}
Since $N_1$ and $N_2$ have no common member in $N$, we can write
\begin{equation}
V=V_1\cup V_2,
\end{equation}
where $V_1\subset N_1\backslash S$ and $V_2\subset N_2\backslash S$ 
are disjoint, and exactly one of them has an odd size. 
and $|V_2|$ is even. 
Now, (\ref{cut of c}) can be rewritten as
\begin{equation}
\label{cut of c 2}
M\triangleq \sum_{i\in V_1} x_i + \sum_{i\in V_2} 
x_i -\sum_{i\in N_1\backslash V_1\backslash S} x_i 
-\sum_{i\in N_2\backslash V_2\backslash S} x_i > |V_1|+|V_2|-1.
\end{equation}
Since $c_1$ and $c_2$ do not generate cuts, all the constraints 
they introduce should be satisfied by $\underline x$. 
Now consider the two sets $V_1\cup S_1$ and $V_2\cup S_1$. 
As $S_1$ is disjoint from both $V_1$ and $V_2$, and exactly 
one of $V_1$ and $V_2$ is odd-sized, we further conclude that 
exactly one of $V_1\cup S_1$ and $V_2\cup S_1$ is odd-sized, 
as well. 
Without loss of generality, assume that $|V_1\cup S_1|$ is odd 
and $|V_2\cup S_1|$ is even.

Depending on whether $|S_f|$ is even or odd, we proceed by 
dividing the problem into two cases:

\emph{Case 1 ( $|S_f|$ is even):} Consider the following constraint 
given by $c_1$
\begin{eqnarray}
\label{const c1}
\sum_{i\in V_1} x_i +\sum_{i\in S_1} x_i + \sum_{i\in S_f} x_i 
-\sum_{i\in N_1\backslash (V_1\cup S_1 \cup S_f)} x_i \leq |V_1|
+|S_1|+|S_f|-1.
\end{eqnarray}
Since $\sum_{i\in S_1} x_i = |S_1|$ and $\sum_{i\in S_0} x_i = 0$, 
we can simplify (\ref{const c1}) as
\begin{eqnarray}
\label{const c1 2}
\sum_{i\in V_1} x_i + \sum_{i\in S_f} x_i - \sum_{i\in N_1\backslash 
(V_1\cup S)} x_i \leq |V_1|+|S_f|-1.
\end{eqnarray}
Now, starting from (\ref{cut of c 2}) and using the fact that 
$N_1\backslash (V_1\cup S) \subset N\backslash V$, we have
\begin{eqnarray}
\label{cut <>}
|V_1|+|V_2|-1 \!\!\!&<&\!\!\! M = \sum_{i\in V_1} x_i + \sum_{i\in V_2} 
x_i - \sum_{i\in N\backslash V} x_i \nonumber \\
\!\!\!&<&\!\!\!\sum_{i\in V_1} x_i + \sum_{i\in V_2} x_i - 
\sum_{i\in N_1\backslash (V_1\cup S)} x_i \nonumber\\
\!\!\!&\leq&\!\!\! |V_1| + \sum_{i\in V_2} x_i - 1 + |S_f| 
- \sum_{i\in S_f} x_i,
\end{eqnarray}
where for the last inequality we used (\ref{const c1 2}). 
Since $\sum_{i\in V_2} x_i \leq |V_2|$, (\ref{cut <>}) yields
\begin{eqnarray}
\label{cut <> 2}
|V_1|+|V_2|-1 < |V_1|+|V_2| - 1 +|S_f| - \sum_{i\in S_f} x_i,
\end{eqnarray}
Hence, $S_f$ is a non-empty set, and since it is even in size, 
we must have $|S_f|\geq 2$.

\emph{Case 2 ( $|S_f|$ is odd):} In this case, we swap the roles 
of $c_1$ and $c_2$, and write for $c_2$
\begin{eqnarray}
\label{const c2}
\sum_{i\in V_2} x_i +\sum_{i\in S_1} x_i + \sum_{i\in S_f} x_i 
- \sum_{i\in N_2\backslash (V_2\cup S_1 \cup S_f)} x_i \leq 
|V_2|+|S_1|+|S_f|-1,
\end{eqnarray}
and for $c_1$
\begin{eqnarray}
\label{const c1, case 2}
\sum_{i\in V_1} x_i +\sum_{i\in S_1} x_i - \sum_{i\in N_1\backslash 
(V_1\cup S_1 \cup S_f)} x_i - \sum_{i\in S_f} x_i \leq |V_1|+|S_1|-1.
\end{eqnarray}
By adding (\ref{const c2}) and (\ref{const c1, case 2}), we obtain
after some cancellation 
\begin{equation}
\label{const c1+c2}
\sum_{i\in V_1} x_i + \sum_{i\in V_2} x_i - \sum_{i\in N_1\backslash 
(V_1\cup S)} x_i - \sum_{i\in N_2\backslash (V_2\cup S)} x_i \leq 
|V_1|+|V_2|+|S_f|-2.
\end{equation}
Note that 
\begin{equation}
\label{subsets...}
N\backslash V = [ N_1\backslash (V_1\cup S) ]\cup [N_2\backslash (V_2\cup S)].
\end{equation}
Hence, combining (\ref{cut of c 2}) and (\ref{const c1+c2}) yields
\begin{equation}
\label{cut <> case 2}
|V_1|+|V_2|-1 < |V_1|+|V_2| +|S_f|-2,
\end{equation}
which means that $|S_f| \geq 2$.
\end{proof}

\section{Proof of Theorem \ref{existence of cycle}}
\label{existence proof}
\begin{proof}
Assume, to the contrary, that for some $i \in \{1, \ldots, K\}$ the cluster
$F_i \subseteq F$ is a tree. Denote by $V=\{v_1,\cdots,v_r\}$ the set of 
indices of variable nodes of the Tanner graph that are in $F_i$, and by 
$V^C$ the complement of this set with respect to $I=\{1,\cdots,n\}$. 
For a vector $\underline u$ of length $n$ and any set
$A \subset I$, let $\underline u_A$ represent the projection of 
$\underline u$ onto the coordinate indices in $A$. Using this notation,
we note that $x_V\in (0,1)^r$.

We now show that there exists a point $\underline{x'}$ in the fundamental 
polytope for which $\underline{x'}_V\in \{0,1\}^r$, 
$\underline{x'}_{V^C}=\underline x_{V^C}$, and 
$\underline\gamma^T \underline{x'} \leq \underline\gamma^T \underline x$.
That is, we identify a point in the fundamental polytope with
cost strictly lower than the cost of $\underline x$, contradicting the
fact that $\underline x$ is a solution to the LP problem.

To accomplish this, we formulate an LP decoding subproblem 
on a Tanner graph $\hat{G}$, which is constructed as follows. 
Let $G'$ be the union of $F_i$ with all the edges and variable 
nodes in $G\backslash F_i$ that were directly connected to a check 
node of $F_i$ in the original graph, $G$. Clearly, since $F_i$ is
disjoint from the other connected components of $F$, any such variable 
node $v$ lies in $G\backslash F$ and therefore has an integer value 
$x_v$ in the solution $\underline x$. 

%


A variable node in $G'\backslash F_i$ might be connected to $k>1$ 
check nodes in $F_i$, in which case $G'$ will contain cycles. 
To break these cycles, we replicate any such variable node, creating
$k$ distinct nodes, each connected to a distinct one of the $k$ check 
nodes. We refer to the resulting graph as $\hat{G}$, and think of
$F_i$ as a subgraph of $\hat{G}$.
We call the variable nodes in $F_i$ and $\hat{G}\backslash F_i$ 
\emph{basic} and \emph{auxiliary} variable nodes, respectively.

To each basic variable node $v$ in $F_i$, we assign the corresponding 
cost $\gamma_v$ in the original LP problem. To each auxiliary 
variable node $\hat{v}$ derived from a parent node $v$ in $G\backslash F_i$, 
we assign a cost $\Gamma$ or $-\Gamma$, according to whether the 
corresponding value $x_v$ in the pseudocodeword $\underline x$ is 
0 or 1.  Here $\Gamma$ is a positive constant that can be chosen,
as described below, to ensure that, in the solution 
$\underline{\hat{x}}$ to the new LP subproblem on $\hat{G}$, the value 
$\hat{x}_{\hat{v}}$ of the auxiliary node $\hat{v}$ is equal to $x_v$.


Since $\hat{G}$ is now an acyclic Tanner graph, the solution, 
$\underline{\hat{x}}$, of LP decoding on this graph will be 
integral \cite{acyclic LP}. On the other hand, if we assign 
to each variable node in $\hat{G}$ the value of the corresponding 
parent node in the original solution, $\underline x$, 
the resulting vector, $\underline{x^*}$, will be another feasible point of the 
LP decoding subproblem on  $\hat{G}$, 
since each check node in  $\hat{G}$  sees the same set of variable node 
values in its neighborhood as does the corresponding check node in $G$
in the original LP decoding problem. 
Furthermore, we claim that if the cost $\Gamma$ is chosen to be larger than 
$\sum\limits_{j \in V} |\gamma_j|$, the values of the auxiliary variable nodes 
will be the same in $\underline{x^*}$ as in the integral optimal solution, 
$\underline{\hat{x}}$.
This follows from the fact that, with the specified assignment of costs 
$\pm \Gamma$ to the auxiliary variable nodes, modifying the value in  
$\underline{x^*}$ of any such node increases the objective function by 
$\Gamma > \sum\limits_{j \in V} |\gamma_j|$. 
No modification of the values corresponding to basic variable nodes can 
compensate for this increase, since any such modification can 
decrease the objective function by at most $\sum\limits_{j \in V} |\gamma_j|$. 
Therefore, auxiliary variable nodes have the same values in both vectors, 
implying that 
\begin{equation}
\label{partial objective}
\sum\limits_{j\in V} \gamma_j \hat{x}_j < \sum\limits_{j\in V} \gamma_j 
{x^*}_j = \sum\limits_{j\in V} \gamma_j x_j.
\end{equation}

We now define the vector $\underline{x'}$ as
\begin{equation}
\label{form x'}
x'_j= 
\begin{cases}
\hat{x}_j \ & \text{if } j\in V,\\
x_j \ & \text{if } j\in V^C.
\end{cases}
\end{equation}
We know that $\underline{x'}_V$ is integral, and (\ref{partial objective}) 
implies that $\underline{x'}$ has a lower cost than $\underline x$. 
It only remains to show that $\underline{x'}$ satisfies all the LP 
constraints introduced by the parity-checks of the original Tanner 
graph $G$. 

Consider a check node $c_j \in G$. If $c_j \in F_i$, its neighboring
variable nodes will have the same set of values as in the solution 
$\underline{\hat{x}}$ of LP decoding on $\hat{G}$, and therefore all 
its corresponding LP constraints will be satisfied. 
If $c_j \notin F_i$, all of its neighboring variable 
nodes are in $V^C$, and they too have the same values as in $\underline x$. 
Hence the LP constraints that $c_j$ introduces will be satisfied. 
It follows that $\underline{x'}$ is a point in the fundamental polytope
with lower cost than $\underline x$, contradicting the fact that 
$\underline x$ is the solution of the original LP decoding problem on $G$.
Therefore, each component $F_i$ must contain a cycle.

\end{proof}

\section*{Acknowledgments}
The authors would like to thank Prof. Marc Fossorier for providing the 
simulation results of the Box-and-Match Algorithm shown in Figs.~
\ref{RPC 32} and \ref{RPC 100}.

\end{document}